\documentclass[twocolumn,showpacs,prb,superscriptaddress,floatfix,amsmath,amssymb]{revtex4}
\usepackage{amsmath}
\usepackage{graphicx}
\usepackage{bm}
\usepackage[usenames]{color}
\usepackage{afterpage}

\def\bra#1{\mathinner{\langle{#1}|}}
\def\ket#1{\mathinner{|{#1}\rangle}}
\newcommand{\braket}[2]{\langle #1|#2\rangle}
\newcommand{\ketbra}[2]{|#1 \rangle \langle #2|}

\newcommand{\commut}[2]{[#1,#2]}
\def\fig#1{Fig.~\ref{#1}}
\def\eq#1{Eq.~\eqref{#1}}

\definecolor{magenta}{cmyk}{0.1,0.8,0,0.1}

\begin{document}
\title{ Coherence loss and recovery of an electron spin coupled
inhomogeneously to a one-dimensional interacting spin bath: an
adaptive t-DMRG study}

\author{Zhi-Hui Wang}
\affiliation{Institute of Theoretical Physics, Chinese Academy of
Sciences, Beijing 100080, China}
\author{Bing-Shen Wang}
\affiliation{State Key Laboratory of Semiconductor Superlattice
and Microstructure and Institute of Semiconductor, Chinese Academy
of Sciences, Beijing 100090, China}
\author{Zhao-Bin Su}
\affiliation{Institute of Theoretical Physics, Chinese Academy of
Sciences, Beijing 100080, China}
\date{\today}


\begin{abstract}

Coherence evolution and echo effect of an electron spin, which is
coupled \emph{inhomogeneously} to an interacting one-dimensional
finite spin bath via hyperfine-type interaction, is studied using
the adaptive time dependent density matrix renormalization group
(t-DMRG) method. It is found that the interplay of the coupling
inhomogeneity and the transverse intra-bath interactions results in
two qualitatively different coherence evolutions, namely, a
coherence preserving evolution characterized by periodic oscillation
and a complete decoherence evolution.  Correspondingly, the echo
effects induced by an electron spin flip at time $\tau$ exhibit
stable recoherence pulse sequence for the periodic evolution and a
single peak at $\sqrt 2 \tau$ for the decoherence evolution,
respectively. With the diagonal intra-bath interaction included, the
specific feature of the periodic regime is kept, while the $\sqrt
2\tau$-type echo effect in the decoherence regime is significantly
affected. To render the experimental verifications possible, the
Hahn echo envelope as a function of $\tau$ is calculated, which
eliminates the inhomogeneous broadening effect and serves for the
identification of the different status of the dynamic coherence
evolution, periodic versus decoherence.

\end{abstract}
\maketitle

\section{Introduction}
Decoherence of a quantum object induced by an interacting spin bath
rather than the conventional boson bath\cite{C-L} is currently an
attractive research subject. 
This interest stems mostly from the belief that electron spin is a
promising candidate of qubit in scalable solid-state quantum
computer architectures.\cite{Kane,Loss} Recently the Hahn echo
technique was also applied to measuring the decoherence of the
localized electron spin. The rapid experimental progress in
single spin measurements promises sensitive measurements of
quantum dephasing effects in the near future.

For an electron spin embedded in a semiconductor matrix, in a proper
range of external magnetic field and at low temperature, the
dominant decoherence source is the temporally fluctuating random
magnetic field originated from the dipolar interaction induced
flip-flops of the surrounding nuclear spin pairs. The electron spin
dephasing due to this random magnetic filed then depends intimately
on the quantum dynamics of the nuclear spin bath. Such an intuitive
dephasing mechanism can be properly described by the hyperfine (HF)
interaction in association with the intra-bath nuclear spin-spin
interactions.

In Ref.\ \onlinecite{Loss2004}, the authors reported a non-Markovian
treatment for the dynamics of a localized electron spin coupled to a
three-dimensional (3D) environment of nuclear spins solely via the
HF interaction. Further work\cite{Das Sarma1,Das Sarma2} stressed
the importance of a quantum dynamic treatment for the intra-bath
spin interactions and developed a quantum cluster expansion method
(see also Ref.\ \onlinecite{Sham3}) to calculate the Hahn echo decay
with its numerical results in good agreement with recent
experiments.\cite{Tyryshkin} Also, in Ref.\ \onlinecite{Sham1}, an
intuitive pseudo-spin model was proposed for the pair interaction in
the nuclear spin bath and predicted a remarkable recoherence effect
at time $\sqrt{2}\tau$ through disentanglement if the localized
electron spin is flipped at time $\tau$.

Almost in parallel, the decoherence of a central spin caused by a
one-dimensional (1D) spin bath is becoming an alternative
interesting subject. The 1D spin bath, e.g., an Ising or XY spin
chain, has the advantage of being analytically solvable or allowing
for more reliable simulations. It was suggested that the decoherence
is subtly related to the characteristic status of the spin bath,
such as correlations, entanglement and criticality.\cite{1d-bath}
For instance, a universal decoherence regime, which is independent
of the system-bath coupling strength, was identified as a
consequence of phase transition. \cite{1d-bath-2}

In this paper we consider an electron spin coupled via HF type
interaction to an \emph{interacting} 1D quantum spin chain, and
employ the adaptive time dependent density matrix renormalization
group (t-DMRG) technique to simulate the underlying quantum
many-body dynamics. 
The model adopted in this work retains the basic ingredients of the
quantum dephasing of an electron spin embedded in a mesoscopic 3D
nuclear spin bath, yet differs from the above mentioned 1D spin-bath
models in that the electron spin is coupled \emph{inhomogeneously}
to all bath spins, where the inhomogeneity is crucial in the
electron spin - bath spin coupling. We show that the interplay of
the HF-type coupling inhomogeneity and the intra-bath interactions
leads to two kinds of qualitatively different behaviors for the
electron spin coherence evolution, namely, fully periodic and
monotonically decaying evolutions, with a crossover regime in
between; And the former corresponds to a coherence preserving phase,
while the latter a complete decoherence phase.

Specifically, (i) The periodic free induction evolution (FIE)
carries a slow modulation with a super-period commensurable to its
basic period.  The competition between the two periods is
responsible for the crossover to the decoherence evolution. In the
decoherence evolution, the coherence decreases monotonically with an
exponential decay index $k=4$. (ii) The echo effects induced by a
sudden electron spin flip show an additional recoherence peak
sequence initiated at $t=\sqrt 2\tau$ or $2\pi-\sqrt 2(\pi-\tau)$ in
the periodic regime, and a single peak at $t=\sqrt 2\tau$ in the
decoherence regime. (iii) The periodic evolution is intrinsic and
stable in sense that, as the chain length increases, the basic
period, the modulation super-period, as well as the corresponding
peak value maintain unchanged, while the peak width keeps shrinking
with an interval of almost zero coherence value emerging from the
midpoint of each basic period. We suggest that, the recoherence
effect can be applied to explore the properties of periodic and
decoherence evolutions with the same almost zero coherence value.
(iv) To render the experimental verifications possible, we show
that, the periodic versus decoherence evolution behaviors can be
unshielded from the inhomogeneous broadening effect by the Hahn echo
envelope at $t=2\tau$.

The rest of this paper is organized as follows:
Sec.~\ref{sec:formulation} is the physical background of our study.
Sec.~\ref{sec:model} is devoted to the motivations, 1D model and
calculation method. Sec.~\ref{sec:results} is the coherence
evolutions in the {\sl single system dynamics} (SSD). Here we denote
SSD\cite{Sham2} as the coherence evolution initiated from any state
of the bath ensemble. Subsec.~\ref{sec:FIE} is for free induction
coherence evolution; Subsec.~\ref{sec:recoherence} is for
recoherence evolution; Subsec.~\ref{sec:Coh_Decoh} is for the
coherence loss (or coherence collapse) phenomena; In
Subsec.~\ref{sec:model_extension}, two variations of the primary
model with modified HF-type interactions are studied.
Sec.~\ref{sec:Hahn_Echo} is devoted to the ensemble average over the
spin bath for the coherence evolution of the electron spin;
Sec.~\ref{sec:conclusion} is discussions and the concluding remarks.
The appendix is devoted to the bath-size dependence of the periodic
evolution.

\section{PHYSICAL BACKGROUND}
\label{sec:formulation}
We consider the quantum dephasing, i.e., the coherence evolution, of
an electron spin embedded in a nuclear spin-$\frac{1}{2}$ bath with
a moderate to strong magnetic field applied. It can be described by
the ensemble average of the off-diagonal matrix element of the
electron spin ${\bf \hat S}_\alpha (\alpha\mbox{=}x,y)$, with the
expression as:
\begin{eqnarray}
\mathcal |\rho_{+-}(t)|=|Tr[\ketbra{-}{+}\hat U^h(t)\hat{\rho}_0\hat
U^{h\dag}(t)]| \;. \label{eq:Coherence0}
\end{eqnarray}
This expression provides itself as a starting point to deal with 
both the free induction coherence evolution and 
the spin echo effect. 
Here, 
$\hat{\rho}_0=\hat{\rho}^e\otimes\hat{\rho}^N$ is the
initial density matrix of the ``electron spin-nuclear spin bath''
system. And $\hat{\rho}^e=\ketbra{\Phi^S}{\Phi^S}$ is the initial
density matrix of the localized spin in a pure state $\ket{\Phi^S}$.
$\hat{\rho}^N=\sum_np_n\ketbra{\Phi^0_n}{\Phi^0_n}$ is that of the
nuclear spin bath where \{$\ket{\Phi^0_n}$\} is a complete set of
the nuclear spin bath and {$\{p_n\}$} is the corresponding
probability distribution.

In \eq{eq:Coherence0}, 
$\hat
U^h(t)=\theta(\tau-t)\hat{U}(t)+\theta(t-\tau)\hat{U}(t-\tau)\hat\sigma_{x,e}\hat{U}(\tau)$
is the evolution operator 
with an instantaneous spin flip imposed to the electron spin at time $\tau$, 
and $\hat{U}(t)=e^{-i\hat{H}t}$ is
the dynamical evolution operator of the ``electron spin - spin bath"
system. The total Hamiltonian
\begin{eqnarray}
\label{eq:Hami} \hat{H}=\ketbra{+}{+}(\hat
H_{Z_n}+\frac{1}{2}\Omega_e+\hat H_+)+\ketbra{-}{-}(\hat
H_{Z_n}-\frac{1}{2}\Omega_e+\hat H_-)
\end{eqnarray}
with $\ket{\pm}$ being the eigenstates of the $z$ component of the
localized spin operator $\bf \hat S$, is decomposed into a direct
sum of two parts, i.e., that projected to the electron spin up
subspace and that to the spin down. And
\begin{eqnarray}
\label{eq:Ham} \hat H_{\pm}=\pm \hat H_0+\hat V_{dd}\pm \hat V_{hf}
\end{eqnarray}
is the effective bath spin Hamiltonian conditioned on the electron
spin polarization.\cite{Das Sarma1,Das Sarma2,Sham1,Sham2}
In \eq{eq:Ham},
\begin{eqnarray}
\hat H_0&=&\frac{1}{2}\sum_iA^{hf}_i\hat I^z_i ,\nonumber\\
 \hat V_{dd}&=&\sum_{i<j}[A^{dd}_{ij}\hat I^z_i\hat I^z_j+\frac{1}{2}B^{dd}_{ij}(\hat I^+_i\hat I^-_j+\hat I^-_i\hat I^+_j)],\nonumber\\
\hat V_{hf}&=&\frac{1}{2}\sum_{i<j}B^{hf}_{ij}(\hat I^+_i\hat
I^-_j+\hat I^-_i\hat I^+_j)
\end{eqnarray}
are the longitudinal hyperfine interaction, intrinsic dipolar
nuclear spin interaction and the effective interaction mediated by
the transverse hyperfine coupling with the electron, respectively.
$\Omega_e$ is the Zeeman energy of the electron spin, while $\hat
H_{Z_n}=\omega_I\sum_i\hat I^z_i$ the Zeeman energy of the nuclear
spin with $\hat I^x_i=\frac{1}{2}(\hat I^+_i+\hat I^-_i), \hat
I^y_i=\frac{1}{2i}(\hat I^+_i-\hat I^-_i)$ and $\hat I^z_i$ being
the nuclear spin operators on the $i$th site.

The coherence can be therefore expressed as
\begin{eqnarray}
    \label{eq:Coherence}
\mathcal |\rho_{+,-}(t)|=|\bra{\Phi^S}\hat S_x-i\hat
S_y\ket{\Phi^S}\otimes\sum_n p_n\braket{\psi^-_n(t)}{\psi^+_n(t)}|
\end{eqnarray}
with
\begin{eqnarray}
    \label{eq:SchodingerP}
\ket{\psi^{\pm}_n(t)}&=&e^{-i\hat H_{Z_n}t}[\theta(\tau-t)e^{\mp\frac{i}{2}\Omega_et}e^{-i\hat H_{\pm}t}\nonumber\\
& &+\theta(t-\tau)e^{\mp\frac{i}{2}\Omega_e(t-2\tau)}e^{-i\hat H_{\pm}(t-\tau)}e^{-iH_{\mp}\tau}]\ket{\Phi^0_n}\;.\nonumber\\
\end{eqnarray}
It shows that the non-Markovian quantum dephasing of an electron
spin is due to the dynamical entanglement between the embedded
spin and the bath (nuclear) spins such that the bifurcated
pathways $\ket{\psi^{\pm}_n(t)}$ detach from each other far apart
as $\braket{\psi^-_n(t)}{\psi^+_n(t)}\rightarrow 0$.
And yet, the coherence can also be restored by the disentanglement
of the localized electron spin from the bath when the bifurcated
pathways of the bath intersect at some later time, i.e.,
$\ket{\psi^+_n(t)}=\ket{\psi^-_n(t)}$.

In the solid-state quantum computer architectures, typical
experiments would be carried out around the temperature $T\sim100
~mK$, which is low temperature for the electron but an extremely high
temperature for the nuclear bath spin dynamics.\cite{Das Sarma1,
Sham2} The ensemble average of the spin bath therefore should be
taken into account to incorporate the extremely low excitation
energy scale of the nuclear spin bath.

Generally, the nuclear spin excitation spectrum of a large system is
the same up to a relative variance $\sim 1/\sqrt N$ ($N$ being the
total number of bath spins)for different initial states $\ket
{\Phi^0_n}$.\cite{Sham2} Thus, the decoherence due to quantum
fluctuations in SSD $|\braket{\psi^-_n(t)}{\psi^+_n(t)}|$ is
insensitive to initial bath state when the system is sufficiently
large and the temperature is appreciable for the nuclear spins.
The coherence \eq{eq:Coherence} can be therefore factorized into
\begin{eqnarray}
\label{eq:factorize}
 |\rho_{+,-}(t)| = \mathcal L_{+,-}(t) \mathcal E_{av}(t) \;,
\end{eqnarray}
where the Loschmidt echo $\mathcal
L_{+,-}(t)\equiv|\braket{\psi^-_n(t)}{\psi^+_n(t)}|$ reveals the
dynamic entanglement between the electron and the nuclear spins in
sense of SSD, and the ensemble average $\mathcal E_{av}(t) \equiv
\sum_n p_n e^{-i\phi_n(t)}$ is the inhomogeneous broadening factor,
with
\begin{eqnarray}
\label{eq:InhomoPhase}
\phi_n(t) &\equiv& -\text{Arg}[\braket{\psi^-_n(t)}{\psi^+_n(t)}]
\nonumber\\
&=& \theta(\tau-t) [\sum_j A^{hf}_j I^{z}_{n,j}(t=0)]~t
\nonumber\\
& & + \theta(t-\tau) [\sum_j A^{hf}_j I^{z}_{n,j}(t=0)]~(t-2\tau) \;.
\end{eqnarray}

Obviously, as shown in \eq{eq:Coherence}, the Zeeman frequency of
the electron spin $\Omega_e$ and nuclear spins $\hat H_{Zn}$ do not
contribute to the coherence, so that the dynamics of the nuclear
spin bath described by the evolution operators $e^{- i\hat
H_{\pm}t}$ maneuvers the coherence of the embedded electron spin.

\section{1D Model: motivations and calculation method} 
\label{sec:model} 

To initiate our discussion, it is interesting to verify that the
diagonal HF interaction contributed $\hat H_0$ can be put into the
form as
\begin{eqnarray}
     \label{eq:H0}
\hat
H_0=\frac{1}{2N}(\sum_{i}A^{hf}_i)\hat{I}^z+\frac{1}{2N}\sum_{i<j}(A^{hf}_i-A^{hf}_j)(\hat
I^z_i-\hat I^z_j)\;.
\end{eqnarray}
Since $\hat{I}^z = \sum_i\hat I^z_i$ commutes with all terms in
$\hat H_\pm$, \eq{eq:H0} shows explicitly that $\hat H_0$ can be
equivalently expressed as a sum of pairwise terms
$(A^{hf}_i-A^{hf}_j)(\hat I^z_i-\hat I^z_j)$ as far as Loschmidt
echo is concerned. On the one hand, the physics of the longitudinal
HF interaction in SSD depends only on the relative differences
$\Delta A^{hf}_{ij}\mbox=A^{hf}_i-A^{hf}_j$, i.e., it is not
affected by the substitution $A^{hf}_j\rightarrow A^{hf}_j+cons.$
for all sites. On the other hand, the dynamic evolution of the spin
bath is then driven by four types of pairwise spin operators $\hat
I^z_i- \hat I^z_j, I ^+_i\hat I^-_j$, $\hat I^-_i\hat I^+_j$ and
$\hat I^z_i\hat I^z_j$. In particular, for $I=\frac{1}{2}$, the
former three form an SU(2) algebra satisfying
\begin{eqnarray}
\begin{cases}
\commut{\frac{1}{2}(\hat I^z_i-\hat I^z_j)}{\hat I^+_i\hat
I^-_j}=\hat I^+_i\hat I^-_j\cr \commut{\frac{1}{2}(\hat I^z_i-\hat
I^z_j)}{\hat I^-_i\hat I^+_j}=-\hat I^-_i\hat I^+_j \cr \commut{\hat
I^+_i\hat I^-_j}{\hat I^-_i\hat I^+_j}=\hat I^z_i-\hat I^z_j
\end{cases} \;,
\end{eqnarray}
while $\hat I^z_i\hat I^z_j$ commutes with the other three terms in
the spin pair subspace. 

Since the 1D model has the advantage of being analytically solvable
or allowing for reliable precise simulations, to set up an 1D model
from its 3D counterpart is often attractive and beneficial. It could
be helpful to clarify some of the essential physics shared with the
corresponding 3D physical system. It might also enrich the
understanding of the quantum dynamics of the decoherence of a
central spin caused by 1D spin bath itself. The observations in the
above paragraph form a guideline in extracting a 1D model from the
decoherence of an electron spin embedded in a 3D mesoscopic nuclear
spin bath.

For a 3D quantum dot, 
$A^{hf}_j$ is proportional to the modulus square of the electron
wavefunction at site $j$, in which the difference $\Delta
A^{hf}_{ij}$ varies significantly site to site.  As the
longitudinal HF coupling plays a major role in the
electron-spin-bath interaction, to keep its distinctive feature of
spatial inhomogeneity in our 1D model, we would like to
take $\Delta A^{hf}_{ij}\mbox=\Delta A^{hf}(j-i)$ 
with $\Delta A^{hf}$ being constant as a minimal model, i.e.,
\begin{eqnarray}
\label{eq:minimal_model} %
A^{hf}_j\mbox{=}(N-j)\Delta A^{hf}\mbox{+}A^{hf}_N \;.
\end{eqnarray}
Moreover, the transverse HF interaction contributed effective 
intra-bath interaction is external magnetic field dependent. 
For a typical semiconductor quantum dot with field strength 
$10-1 ~T$,\cite{Das Sarma2, Sham2}
it has an energy scale of $B^{hf}_{ij}\sim1-10~s^{-1}$. 
Meanwhile that of the near-neighbor dipolar intra-bath interaction 
is $B^{dd}_{ij} \sim 10^2~s^{-1}$. 
The former is much smaller than that of the longitudinal part by 
an order of $A^{hf}_i/\Omega_e$, it is also smaller than 
the latter by an order of 10.
We therefore neglect the extrinsic nuclear spin interaction and set
$B^{hf}_{ij}\mbox{=}0$. We further take the intrinsic dipolar
interaction to be nearest-neighbored as
$B^{dd}_{ij}\mbox{=}B^{dd}\delta_{i,j\pm1}$, and
$A^{dd}_{ij}\mbox{=}A^{dd}\delta_{i,j\pm1}$.
Then the electron-spin conditioned effective bath Hamiltonian becomes
\begin{eqnarray}
\hat H_{\pm}&=&\frac{\Delta A^{hf}}{2} \Big(\mp\sum^N_{j=1}j\hat
I^z_j
+ \frac{1}{\xi} [\sum^{N-1}_{j=1} -(\hat I^+_j\hat I^-_{j+1}+\hat I^-_j\hat I^+_{j+1})
\nonumber\\
& &
+2\eta \hat I^z_j\hat I^z_{j+1}]\Big)
\pm\Big(\frac{A^{hf}_N+N\Delta A^{hf}}{2}\Big)\hat I^z \;.
\end{eqnarray}
Now the dynamic evolution of the quantum coherence
can be characterized by dimensionless parameters
$\xi\equiv {\Delta A^{hf}}/(-{B^{dd})}$
and $\eta\equiv {A^{dd}}/(-{B^{dd})}$,
with the time variable $t$ normalized by
$4/\Delta A^{hf}$ (throughout the paper). 
In sum, such extracted 1D system retains the essential features of
its 3D counterpart, e.g., the monotonically descending spatial
inhomogeneity of the longitudinal HF interaction, the pairwise SU(2)
algebra and that the sign of $\hat H_0$ is conditioned on the
embedded electron spin state. The t-DMRG method\cite{Daley} is
available and can be properly applied to simulate the time evolution
with enough precision.

In our t-DMRG calculations, the chain length is taken as
$N\mbox{=}50, 100, 150, 200 $ and $250$, with properly chosen
hundreds of truncation states being kept. Since, as shown in
Appendix \ref{app:chain_length}, the coherence evolution is
stablized after $N\gtrsim 30$, the calculations in our manuscript
are proceeded with $N=50$ if not specified. The Trotter-Suzuki
decomposition\cite{Suzuki} is held up to the 2nd order. If
$\xi\mbox=100$, the time step is chosen as $\Delta t \sim 10^{-2}$,
the numerical precision can be maintained for $10^4$ time steps. Our
initial bath states are prepared as the eigenstates of the Zeeman
energy of the bath by an independent DMRG calculation,\cite{Gobert}
where off-diagonal coherence is absent.


\section{Single-system dynamics} 
\label{sec:results} 
\subsection{Free induction coherence evolution in single-system dynamics}
\label{sec:FIE} 
We initiate our discussion from the single state coherence evolution,
in which the initial bath spin state is taken as any of
the eigenstates of the Zeeman energy of the bath spin.

For simplicity, we first take the longitudinal intra-bath interaction
as $A^{dd}=0$.
We find that the coherence evolution is controlled by two kinds
of distinct behaviors, the periodic evolution in the regime with
$\xi$ large enough and the decoherence evolution with $\xi$
small enough.  The gradient of the longitudinal HF interaction
$\Delta A^{hf}$ favors the periodic evolution while the intra-bath
spin interaction causes decoherence.

\begin{figure}[htbp]
\begin{center}
    \includegraphics[width=3in]{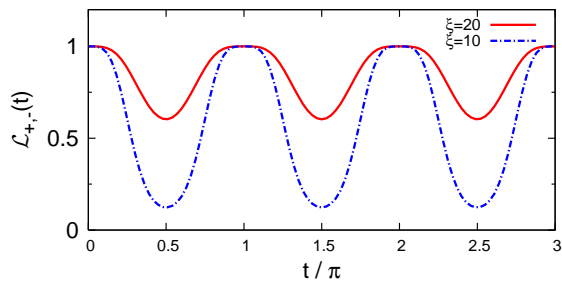}
\end{center}
 \caption[FIE_PR]
{ (color online). Free induction coherence evolution in the periodic
regime with $\xi=10,~20$, $A^{dd}=0$. The time is normalized by
$4/\Delta A^{hf}$. The evolution exhibits a basic periodic $\pi$.
 } \label {FIE_PR}
\end{figure}

\begin{figure}[htbp]
\begin{center}
\includegraphics[width=3in]{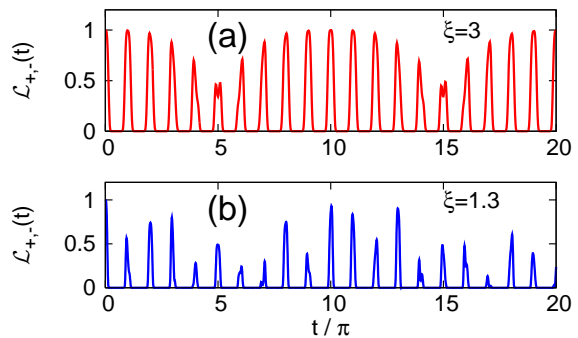}
\end{center}
 \caption[FIE_CR]
{(color online).
Crossover behavior of free induction coherence evolution with $A^{dd}=0$.
(a) For $\xi=3$, the periodic evolution carries a
modulation with period $10 T_{dd}$.
(b) For $\xi=1.3$, the peaks at
$t = m T_{dd},~m=1,2\ldots$ oscillate irregularly.
}
\label {FIE_CR}
\end{figure}

\begin{figure}[htbp]
\begin{center}
\includegraphics[width=3in]{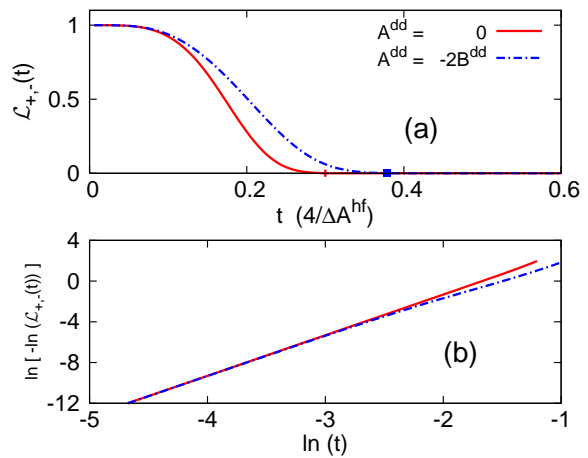}
\caption[FIE_DR]
{
(color online).
(a) Free induction coherence evolution in the decoherence regime
for $\xi=0.5$, with $A^{dd}=0$ (red solid line) and $A^{dd}=-2B^{dd}$ (blue dashed line).
the coherence decreases monotonically. Coherence disappearing time $t_d$ is
labeled on the $t$ axis.
(b) Logarithmic plot for (a).
The exponential decay indices are $k\mbox=4$ for both evolutions.
}
\label {FIE_DR}
\end{center}
\end{figure}

For $\xi \gg 1$,
see $\xi\mbox=10$ in Fig.~\ref{FIE_PR}, the
coherence evolution exhibits a periodic oscillation versus the
reduced time with a generic period $T_{dd}\cong\pi$ independent of $\xi$.
It shows also a mirror symmetry with respect to the midpoint of each period.

Moreover, if we attend to the time scale of the order of $\xi~ (\gg
T_{dd})$, we find that the oscillating evolution carries a slow
modulation with a super-period $T_{sup}$ locked at an integer multiplier of
$T_{dd}$ as
\begin{equation}
\label{eq:T_sup}
T_{sup}\cong
(1+[\xi^2]^{fl,ce})T_{dd}\;,
\end{equation}
where the superscript ``$fl$" means the floor function 
and ``$ce$" the ceiling function. 
For any real number $x$, we denote the integer closest to $x$ by $[x]^{fl,ce}$.
The coexistence of two mutually commensurate periodicities is
essentially the consequence of two kinds of energy scales as that of
the HF interaction and that of the intra-bath interaction. When
$\xi$ is large enough ($\Delta A^{hf}\gg B^{dd}$), such slow
modulation does not affect the above discussions which mainly extend
only a few $T_{dd}$ periods. When $\xi$ gets down to the lower brim
of the periodic regime, as a precursor of the crossover zone, the
modulation becomes palpable and is then responsible for the
evolution periodicity, see \fig{FIE_CR} (a).

With the further decrease of $\xi$, the super-period becomes
comparable to $T_{dd}$, and its periodicity starts to disappear,
i.e., its peak sequence starts to decay in a chaotic way, see
\fig{FIE_CR} (b). The time scale of the basic period $T_{dd}$
appears quite robust such that its remnant persists even up to the
peripheric regime of the decoherence evolution.

When $\xi$ is small enough, the coherence attenuates monotonically
approaching zero within a time interval $t_d$ and will not rise up
perceptibly any more, see Fig.~\ref{FIE_DR} (a). The decoherence time
evolution can be fitted as an exponential function of minus $t$ to
the power $k$ with $k\mbox=4$.  see Fig.~\ref{FIE_DR} (b).  Such
decoherence behaviors in our 1D model are in excellent consistence
with those deducted from the pseudo-spin model in 3D\cite{Sham2}.

Now we further introduce the diagonal spin-spin interaction with
$A^{dd}=-2B^{dd}$, ($\eta = 2$), which is consistent with the nuclear intrinsic
dipole-dipole interaction.\cite{Das Sarma2}
The coherence evolution still exhibits two types of qualitatively different
behaviors, i.e., periodic oscillation and complete decoherence evolution.

\begin{figure}[htbp]
\begin{center}
    \includegraphics[width=3in]{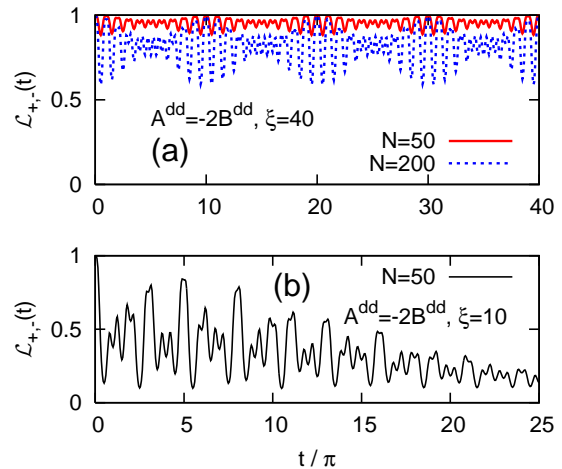}
\end{center}
 \caption[Effect_L_Add]
{ (color online).
Free induction coherence evolution with $A^{dd}=-2B^{dd}$.
(a) Modulation in the periodic regime with $\xi=40$, for $N\mbox=50,~200$.
For $A^{dd}\mbox=0$ (not shown here), the modulation on the periodic evolution
is negligible (see \eq{eq:T_sup}) while here, with $A^{dd}$ included,
the modulation is strengthened to $T_{sup}\cong 20 T_{dd}$.
The super-period $T_{sup}$ is not affected by chain length.
(b) For $N=50$, $\xi=10$, the evolution enters the crossover regime already.
}
\label {Effect_L_Add}
\end{figure}

In the periodic regime,
although the diagonal interaction constant $A^{dd}$
is of the same order of magnitude with that of $B^{dd}$,
it makes the modulation considerably strengthened
with the modulation period shortened and its amplitude enhanced.
Instead of $([\xi^2]^{fl,ce}+1) T_{dd}$ as in the case of $A^{dd}=0$,
the modulation period becomes much smaller as
\begin{equation}
T_{sup}=[\frac{\xi}{2}]^{fl,ce} T_{dd}\;,
\end{equation}
see \fig{Effect_L_Add} (a).

The inclusion of the diagonal intra-bath interaction encroaches the
periodic regime of the case with $A^{dd}=0$ and extends the
crossover regime. Take $\xi=10$ as an example, instead of exhibiting
typical periodic behavior with $A^{dd}=0$, the peak sequence of the
coherence evolution decays in a chaotic way with $A^{dd}=-2B^{dd}$
included, see \fig{Effect_L_Add} (b).

In the decoherence regime, the
coherence evolution behavior is very close to that with $A^{dd}=0$,
it keeps not only almost same coherence disappearing time $t_d$,
but also the same decay index $k=4$, see \fig{FIE_DR}.

\subsection{Recoherence effects in single-system dynamics} 
\label{sec:recoherence} 
\begin{figure}[htbp]
\begin{center}
    \includegraphics[width=3in]{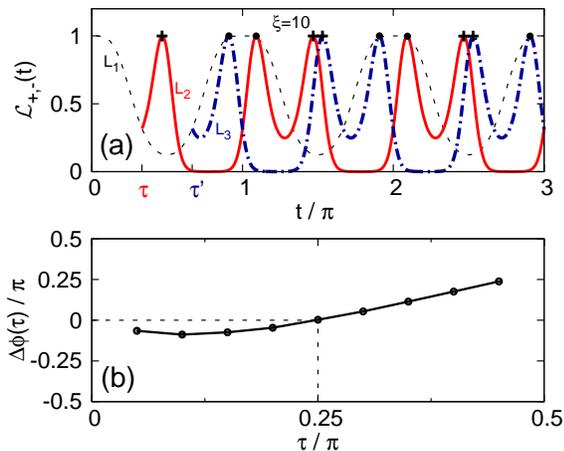}
\end{center}
 \caption[Recoh_PR]
{ (color online). Recoherence behavior in the periodic regime with
$\xi=10$. (a) Coherence evolutions without spin flip ($L_1$), and
with spin flip at $\tau\mbox=T_{dd}/3$ ($L_2$), and
$\tau'\mbox=T_{dd}-\tau$ ($L_3$). The recoherence peaks and
intervened original peaks are labeled on the lines as crosses and
filled circles, respectively. After $t>\tau'$, $L_2$ and $L_3$ show
a mirror symmetry with respect to reflection points $t=n\pi/2$ with
$n=2,3,\ldots$ (b) The phase shift $\Delta \phi$ as a function of
$\tau$ with $\tau\in(0,T_{dd}/2)$. $\Delta \phi(\tau)$ changes sign
at $t\mbox={T_{dd}}/{4}$, at which $\Delta\phi\mbox=0$. } \label
{Recoh_PR}
 \end{figure}

We then impose a sudden electron spin flip to the coherence
evolution at $t=\tau$ and consider the case $A^{dd}=0$ first.
For the periodic evolution, with
$\tau\in(0,{T_{dd}}/{2})$, it would yield an additional periodic
pulse sequence at $\sqrt 2\tau\mbox+nT_{dd}$ with $n\mbox=0,1\ldots$
Meanwhile, those coherence peaks in the absence of spin flip still
survive but, due to the intervention of the spin flip, experience a
phase delay $\Delta\phi(\tau)$ and shift to
$t\mbox=\Delta\phi(\tau)\mbox+mT_{dd}$ with $m\mbox=1,2, \ldots$,
see Fig.~\ref{Recoh_PR}. 
For $\tau\in({T_{dd}}/{2},T_{dd})$, the recoherence evolution is
found to be the reflection image of that resulting from spin flip at
$T_{dd}-\tau$, so that the recoherence peak sequence initiates at
$2T_{dd}-\sqrt 2(T_{dd}-\tau)$ and $\Delta \phi(\tau)=-\Delta
\phi(T_{dd}-\tau)$, see Fig.~\ref{Recoh_PR}(a). This behavior is a
direct consequence of the mirror symmetry of the free induction
periodic evolution, with respect to reflection points
$(n\mbox+{1}/{2})T_{dd}$.
We further realize that, as expected, the recoherence evolution
exhibits the same super-period as that of the free induction
evolution.

\begin{figure}[htbp]
\begin{center}
    \includegraphics[width=3in]{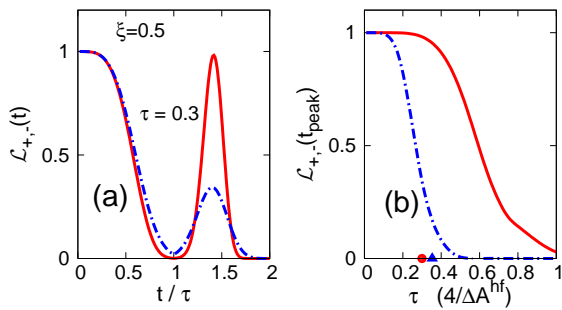}
\end{center}
 \caption[Recoh_DR]
{ Recoherence evolution in the decoherence regime for $\xi=0.5$,
$A^{dd}=0$ (red solid lines), and $A^{dd}=-2B^{dd}$(blue dashed
lines). (a) The coherence peaks at $\sqrt 2 \tau$. (b) The peak
value $\mathcal L_{+,-}(t_{peak})$ as a functions of $\tau$.  The
coherence disappearing time $t_d$'s are labeled on the $\tau$ axis
in the same color with the corresponding curves. } \label {Recoh_DR}
\end{figure}

In the decoherence regime, the recoherence signal closely depends on
the time $\tau$ when the spin flip is imposed. As long as
$\tau<t_d$, the recoherence pulse will peak at
$t\mbox{=}\sqrt{2}\tau$ with the coherence fully recovered as shown
in \fig{Recoh_DR} (a). However, if the spin flip time $\tau$ exceeds
$t_d$, the coherence recoveries still at $\sqrt{2}\tau$ but damps
quickly, see solid lines in \fig{Recoh_DR} (b).

With the longitudinal spin-spin interaction included $A^{dd}=-2B^{dd}$,
in the periodic regime, the recoherence behavior exhibits similar evolution to that with $A^{dd}=0$.
The recoherence peak sequence experiences the same enhanced
modulation as in the case without spin flip. In the decoherence
regime, however, even with the spin flip exerted within $t_d$, with
increase $\tau$, the peak value at $\sqrt 2 \tau$ decreases
considerably faster than that without $A^{dd}$, as shown in
\fig{Recoh_DR} by the dashed lines.

To our understanding, the above calculated $\sqrt
2\tau$-type spin echo effect validates the intuitive
mechanism of the recoherence effect proposed in Ref.\
\onlinecite{Sham2} for a mesoscopic 3D quantum dot. Yet the diagonal
intra-bath interactions considerably weakens the
recoherence effect in the decoherence evolution in 1D case.

\subsection{Coherence loss versus coherence preserving}
\label{sec:Coh_Decoh} 

\begin{figure}[htbp]
\begin{center}
    \includegraphics[width=3in]{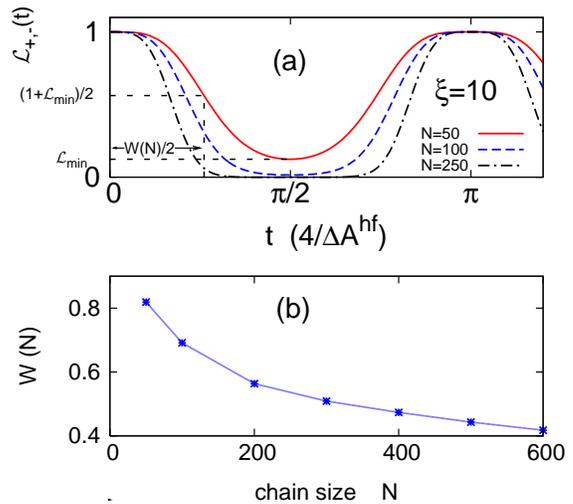}
\end{center}
 \caption[Effect_L_Pi]
{ (color online). Effect of chain size N on the basic periodicity in the periodic regime.
$\mathcal L_{+,-}(t)$.
(a) For $\xi=10$, $A^{dd}\mbox=0$,
when $N$ varies from $50$ to $250$,
the oscillating period $T_{dd}$ is independent of chain size.
$W(N)$ labels the width of the coherence peak.
(b) The width of the peak $W(N)$ defined in (a)
as a function of the bath chain size.
}
\label {Effect_L_Pi}
\end{figure}

We have studied the bath size dependence of the coherence evolution
in detail starting from $N=2$, see Appendix \ref{app:chain_length}.
We realize that, when the chain size $N$ exceeds $30$ sites, the
coherence evolution in the periodic regime becomes stable in sense
that, the basic period, the modulation super-period, as well as the
corresponding peak value maintain unchanged; Meanwhile, the dip of
each basic period moves down gradually, and the width of the peak
keeps shrinking as the chain length increases, see
\fig{Effect_L_Pi}.

Such shrinking of the peak width of the periodic coherence evolution
is essentially a distinctive feature sharing with the Hepp-Coleman
type models\cite{Hepp, Bell, Namiki, Kobayashi, Nakazato, Sun_1993}.
In particular, after the coherence at the dip of each period drops
down to a value close to zero, an interval with almost zero
coherence value emerges from the dip and stretches to both sides
within each period. In fact, in the Hepp-Coleman type models, the
evolution operator can be factorized into a continued multiplication
over chain sites due to the non-interacting character of the spin
bath. What we have here is an interacting spin bath with its
evolution operator cannot be factorized.

\begin{figure}[htbp]
\begin{center}
    \includegraphics[width=3in]{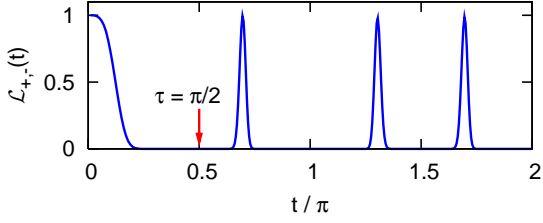}
\end{center}
 \caption[L200_Recoh]
{ (color online). Recoherence evolution for $N=200$, $\xi=5$,
The spin flip is exerted when the coherence value is almost zero.
}
\label {L200_Recoh}
 \end{figure}

Such almost-zero coherence resulted from the successive separation
of the bifurcated pathways was often regarded as an indication of
the loss of coherence.\cite{Bell, Sun_2001} However, if a spin flip
is imposed to the electron at time $\tau$ within the almost-zero
coherence interval for the periodic evolution, the spin flip will
always induce a peak sequence with its coherence value fully
recovered, see \fig{L200_Recoh}. Yet, in the decoherence regime, if
the spin flip is exerted at a time when the coherence value is as
small as those in the zero-coherence interval, no prominent
recoherence signal can be induced.

We therefore have the understanding that the periodic evolution is
coherence preserving in sense that the spin-flip operation always
induces a stable recoherence peak sequence, while the decoherence
evolution with the coherence monotonically decreasing is actually a
coherence-losing process. Our model accommodates these two
qualitatively different coherence evolutions connected by a
continuous crossover, and under the control of the competition
between the electron spin - bath interaction and the intra-bath
interaction.

\subsection{Extension of the minimal model}
\label{sec:model_extension} 

Besides the study in the above subsections, it is also interesting
to examine the effect of relaxing the condition of constant $\Delta
A^{hf}_{j,j+1}$, i.e., \eq{eq:minimal_model} in our minimal model.

\begin{figure}[htbp]
\begin{center}
    \includegraphics[width=3in]{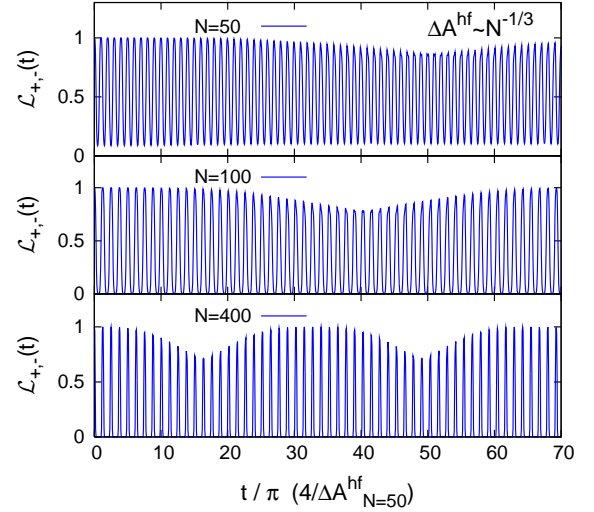}
\end{center}
 \caption[Vary_dAhf]
{ (color online).
Coherence evolution in the periodic regime with
$\Delta A^{hf}\propto N^{-1/3}$,
with chain size $N=50,100,400$.
Time unit is taken as $\Delta A^{hf}$ when $N=50$.
The effective $\xi$ equals $10,~8,~5$, respectively.
Accordingly, the super-period shrinks as the chain size increases.
}
\label {Vary_dAhf}
 \end{figure}

We first consider the case with $\Delta A^{hf}$ being chain size
dependent $\Delta A^{hf}\propto N^{-1/3}$ but maintaining itself as
a site-independent constant. In Appendix \ref{app:chain_length}, we
have numerically verified that the coherence evolution is stable
with the increase of bath chain. In the present case, as chain size
increases, the slope of the hyperfine interaction decreases. We
calculated the coherence evolution for $N=50,100,400$, with $\Delta
A^{hf}= (50/N)^{1/3} $, and fixed $B^{dd}=-0.1$. The results are shown in
\fig{Vary_dAhf}. For $N=50$, the evolution parameter is $\xi=10$,
the coherence evolution exhibits typical periodic behavior. As bath
size increases, the super-period of the periodic evolution shrinks,
since the effective $\xi$ decreases as chain length extends.
Therefore, it can be understood that,
as the chain size increases,
due to the decreasing of the inhomogeneity of the HF interaction,
the system shows up a tendency from periodic oscillation
to crossover into a decoherence evolution.

\begin{figure}[htbp]
\begin{center}
    \includegraphics[width=3in]{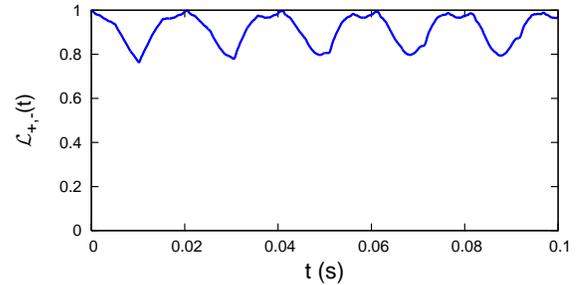}
\end{center}
 \caption[Cos_Ahf_PR]
{ (color online). Coherence evolution with $A^{hf}_j\mbox=\tilde
A^{hf} \cos^2 \frac{(j-1)\pi}{2N}$ in the coherence-preserving regime with
$B^{dd}\mbox=100 ~s^{-1}$, and $A^{dd}\mbox=0$.
For $N=200$, $\tilde A^{hf}=2\times10^7~s^{-1}$,
the coherence oscillates with its value close to unity.
}
\label {Cos_Ahf_PR}
 \end{figure}

\begin{figure}[htbp]
\begin{center}
    \includegraphics[width=3in]{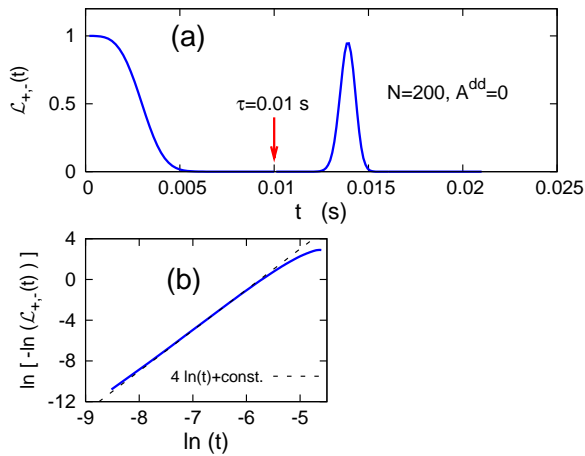}
\end{center}
 \caption[Cos_Ahf_DR]
{ (color online). For $N=200$ with $A^{hf}_j\mbox=\tilde
A^{hf} \cos^2 \frac{(j-1)\pi}{2N}$, $\tilde A^{hf}\mbox=10^5~s^{-1}$,
$B^{dd}\mbox=100 ~s^{-1}$, and
$A^{dd}\mbox=0$.
(a) recoherence evolutions behaves similar to that in the
decoherence regime of our primary model.
(b) Logarithmic plot of coherence evolution in (a) before spin flip,
the evolution also exhibits quartic exponential coherence decay.
}
\label {Cos_Ahf_DR}
 \end{figure}

We now consider the effect of the site-dependence of
$A^{hf}_{j,j+1}$. Along certain crystalline symmetry axis,\cite {Das
Sarma3} the wave function of the embedded electron in a 3D quantum
dot has the form of a cosine function. We take $A^{hf}_j\mbox=\tilde
A^{hf} \cos^2 \frac{(j-1)\pi}{2N}$,
Then, $\Delta A^{hf}_{j,j+1}$ varies from site to site on the lattice
chain. and the average slope of $A^{hf}_j$ is chain-size dependent
for a given $\tilde A^{hf}$.
Take the chain length $N=200$ and $B^{dd}\mbox=10^2 ~ s^{-1}$.
If $\tilde A^{hf}$ is set as $2*10^7~s^{-1}$,
we find that, the periodicity disappears due to the lack of a
unified time normalization $\Delta A^{hf}$,
although the site-dependent ${\Delta A^{hf}_{j,j+1}}/{B^{dd}}$
is of the order $200$,
which falls into the periodic regime of our primary model.
Nevertheless, interestingly, the non-periodic
coherence evolution keeps oscillating close to unity without
decaying, see \fig{Cos_Ahf_PR}. For $\tilde A^{hf}=10^5~s^{-1}$,
${\Delta A^{hf}_{j,j+1}}/{B^{dd}}$ is of the order $1$,
the coherence and recoherence evolutions behave almost the same way as their
counterparts in decoherence regime of our primary model, see
\fig{Cos_Ahf_DR}. We stress that, the larger chain size $N$,
the larger $\tilde A^{hf}$ is required so stay in the coherence
preserving regime.
for any chain size $N$
(at least within our calculation experience), the non-decaying
coherence preserving regime always exists
if we choose $\tilde A^{hf}$ large enough,

In fact, in a 3D quantum dot, the wavefunction of the embedded electron
extends over the whole dot with a normalization factor inversely
proportional to the square root of the dot volume. The longitudinal
HF interaction $A^{hf}_j$, which is proportional to the modulus
square of the electron wavefunction at site $j$,\cite{Feher} is then
not only dependent on the specific form of the wavefunction site by
site, but also dot-volume dependent. The above two variations on the
HF-type interaction in our 1D model resembles these two aspects of the 3D
quantum dot.

The discussions in this section imply that,
for a general class of inhomogeneous interaction,
the dynamic coherence evolution is still under the control of two
fix points, i.e., a decoherence regime with its whole feature kept
as in our primary model, and corresponding to the periodic regime, a
quantum chaotic regime instead, in which the
coherence sustains 
a non-decaying irregular oscillation.

\section{Inhomogeneous broadening and Hahn echo envelop} 
\label{sec:Hahn_Echo} 
The single system dynamics studied in the above section exhibits
various important features of the quantum dynamic entanglement process.
Yet, these interesting properties could be shielded by the inhomogeneous
broadening effect contributed by the ensemble average ${\cal E}_{av}(t)$.
It is then desirable to limit the bath spin configurations experimentally
to observe the SSD.

\begin{figure}[htbp]
\begin{center}
    \includegraphics[width=3in]{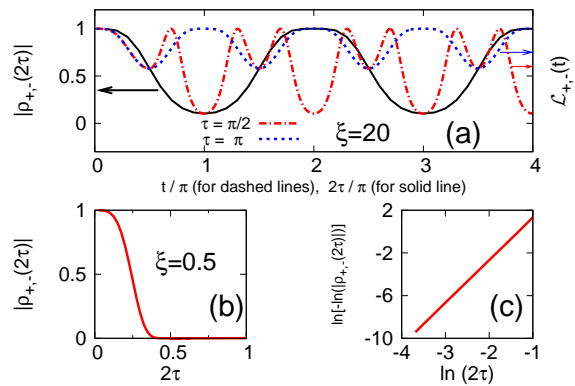}
\end{center}
 \caption[HahnEcho_Sig]
{ (color online). Hahn echo envelope $|\rho_{+,-}(2\tau)|$ as
functions of $2\tau$ with $A^{dd}\mbox=0$. High temperature
approximation for the nuclear spin bath ensemble is taken as
$p_n=\text{const.}$\cite{Das Sarma1,Sham2} (a) For $\xi=20$,
$|\rho_{+,-}(2\tau)|$ (solid line) is a periodic function with
period $2T_{dd}$. The short dashed line is the recoherence evolution
in SSD for $\tau \mbox= T_{dd}$, where the spin flip effects
virtually disappears; it intersects the Hahn echo envelope at
$t\mbox=2nT_{dd},~n=1,2\ldots$ The long dashed line is for $\tau
\mbox= {T_{dd}}/{2}$, and intersects the Hahn echo envelope at
$nT_{dd}$. (b) and (c) show that for $\xi\mbox=0.5$, the Hahn echo
signal exhibits a quartic exponential decay.
}
\label {HahnEcho_Sig}
 \end{figure}

Nontheless, if an electron spin flip is imposed at time $\tau$,
the broadening effect due to phase accumulation
will be canceled
as $\mathcal E_{av}(2\tau)\mbox=1$ at $t\mbox=2\tau$,
see \eq{eq:InhomoPhase}.
The coherence
$|\rho_{+,-}(2\tau)|$ then results solely from the dynamic quantum
entanglement as
\begin{eqnarray}
|\rho_{+,-}(2\tau)|\mbox=\mathcal L_{+,-}(2\tau)
\end{eqnarray}
and serves an exposure of the quantum coherence evolution in SSD. We
plot $|\rho_{+,-}(2\tau)|$ as functions of $\tau$ in \fig{HahnEcho_Sig}.
In fact, in the periodic regime,
compare the recoherence evolutions induced by
spin flip exerted at $\tau$ and $\tau+T_{dd}$ respectively,
since the coherence status at $\tau$ and $\tau+T_{dd}$ are the same,
meanwhile the recoherence evolutions are also periodic with period $T_{dd}$,
we realize that the Hahn echo signal as a function of $2\tau$ is
periodic with a period $2T_{dd}$.
We further provide two examples to illustrate
its physical implications, see dashed lines in \fig{HahnEcho_Sig}.
In the decoherence regime, the Hahn echo
exhibits a quartic exponential decay as shown in \fig{HahnEcho_Sig} (b),
consistent with that reported in 3D case.\cite{Sham2}
Therefore, in virture of the Hahn echo technique,
we can detect the two qualitatively different
dynamic coherence evolutions in SSD.

We notice that, if $A^{hf}_N$ can be adjusted to be
$A^{hf}_N \mbox= m \Delta A^{hf},~m\mbox=0,1,...$, the inhomogeneous
broadening factor $\mathcal E_{av}(t)$ will be refocused to $1$ at
$t\mbox=m{T_{dd}}/{2}$, which would provide directly exposures of
coherence evolution in SSD without spin flips.

\section{concluding remarks}
\label{sec:conclusion} 
In this paper, we studied the coherence evolution of an electron
spin interacting with a 1D finite spin chain via inhomogeneous
hyperfine-type coupling. This model is extracted from its 3D
counterpart: an electron-spin dephases due to the interaction with
the nuclear spins in a mesoscopic 3D quantum dot.

We realized that in our model, the dynamic evolution of the electron
spin coherence manifests a continuous crossover from a periodic to a
decoherence evolution within our numerical precision.
The periodic evolution carries a modulation with a commensurate
super-period, and the competition between the two periods is
responsible for the crossover. The counterpart of the periodic
evolution in our model was not found in the 3D case.
The decoherence evolution as well as its $\sqrt 2 \tau$-type echo
effect validates most of the distinctive features resulted from the
pseudo-spin model or cluster-expansion method for a 3D quantum dot.

We included the longitudinal intra-bath spin-spin interaction.
Physically, it is a kind of scattering term between the collective
excitations, and provides a decoherence-favorable mechanism. We
found that, the system still exhibits periodic evolution as long as
the deep inhomogeneity of HF-type interaction dominates. In the
decoherence regime, the longitudinal interaction has little effect
on the free induction evolution, but would significantly affect the
$\sqrt 2 \tau$-type echo effect.

As the bath size increases, the periodic evolution shows up a ``loss
of coherence" phenomena, i.e., an interval with almost zero
coherence value emerges from the midpoint and stretches to both
sides within each evolution period. Nevertheless, by imposing the
spin echo effect, we learn that the periodic evolution is
essentially a coherence-preserving evolvement while the decoherence
evolution is a coherence losing process.

We investigated two variations of our primary model with modified
hyperfine-type interaction. The dynamic coherence evolution still
exhibits two kinds of qualitatively different behaviors.

In case the proposed 1D system can be realized experimentally, e.g.,
by applying the optical lattice manipulation techniques to cold atom
system,\cite{Morsch} the Hahn echo type refocusing effect could then
unveil the rich phenomena in single-system dynamics.

As the final remarks, we would like to stress that, increase the
gradient of hyperfine interaction $\Delta A^{hf}_{j,j+1}$ favors to
keep the coherence evolution in the periodic regime. Therefore,
steepening the inhomogeneity of the wavefunction of the embedded
electron could be favorable to preserve the coherence. This might
shed light on the materialization of prolonging the quantum
coherence time of the embedded electron spin. Moreover, for the
evolution status with the same almost-zero coherence values,
the different responses to the electron spin flip
suggests an unexplored and rather intriguing aspect
of the coherence description for the electron spin interacting
with a nuclear spin bath.

\acknowledgments
 We are grateful to Prof. Chang-Pu Sun for bringing
our attentions to this interesting topic, to Prof. Xin-Qi Li
and Dr. Hui Tang for helpful discussions.

\appendix
\section{Chain length dependence of the periodic evolution}
\label{app:chain_length}
It is crucial to verify that in the regime of $\xi\gg1$, the
periodic coherence evolution is stable with the increase of the
chain length.

\begin{figure}[htbp]
\begin{center}
    \includegraphics[width=3in]{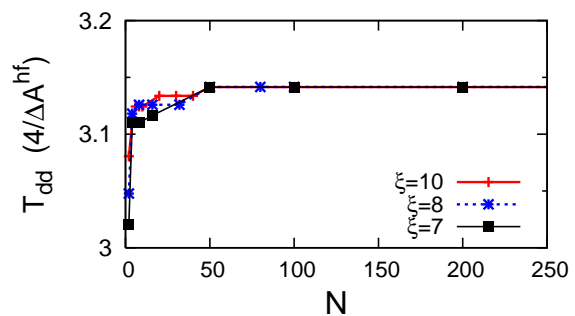}
\end{center}
 \caption[Tdd_L]
{ (color online).
Basic period $T_{dd}$ as a function of chain size $N$
for $\xi=10,~8,~7$, with $A^{dd}=0$.
The initial configuration of the bath spin is randomly picked.
}
\label {Tdd_L}
\end{figure}

\begin{figure}[htbp]
\begin{center}
    \includegraphics[width=3in]{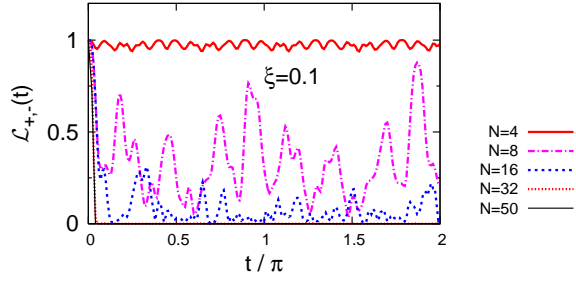}
\end{center}
 \caption[PR_to_DR]
{ (color online).
Coherence evolution for different bath size
with $\xi=0.1$ and $A^{dd}=0$.
The decoherence evolution comes into being for $N>30$.
}
\label {PR_to_DR}
\end{figure}

We first consider the chain length dependence of the periodic
evolution for N being small. For convenience, we take $A^{dd}=0$,
and focus on the first few periods. We have exact solution for the
electron spin interacting with a 2-sites nuclear spin bath, the
solved basic period is $T_{dd}(N=2)=\pi/\sqrt {1+4/\xi^2}$. As the
chain size increases from $N=2$, our numerical calculations show
that, see \fig{Tdd_L}, the basic period increases correspondingly
and stabilizes at a value $\pi$ for $N\gtrsim 30$ approximately,
which is independent of $\xi$. And the super-period is also stable
for $N>2$. For comparison, we consider additionally the coherence
evolution beyond the periodic regime with, e.g., $\xi=0.1$. As shown
in \fig{PR_to_DR}, for $N\gtrsim 30$, the decoherence evolution
comes into appearing as a monotonically decaying evolution with an
exponential decay index $k=4$. The corresponding calculation
proceeds over $100\pi$ without any coherence revival effect. These
two mutually complemented aspects indicate that the coherence
evolution is indeed stable for chain length $N\gtrsim 30$. The
calculated various properties of the coherence evolution for
$N\geq50$ are intrinsic.

\begin{figure}[htbp]
\begin{center}
    \includegraphics[width=3in]{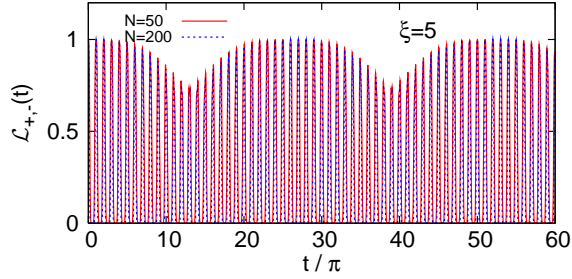}
\end{center}
 \caption[Effect_L]
{ (color online).
Modulation on the periodic evolution.
for $N=50,~250$ with $\xi=5$, $A^{dd}=0$.
The super-period is not affected by chain length.
}
\label {Effect_L}
\end{figure}

\begin{figure}[htbp]
\begin{center}
    \includegraphics[width=3in]{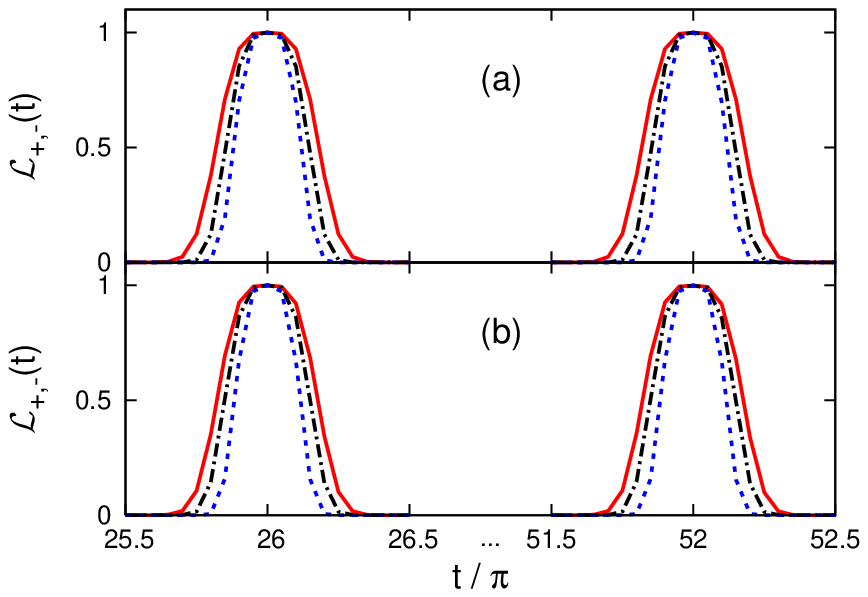}
\end{center}
 \caption[Sup_Peak]
{ (color online).
Modulation on the periodic evolution with $A^{dd}=0$ for $\xi=5$.
The first two peaks of the super modulation at $t=T_{sup}$ and $t=2T_{sup}$
are shown in detail.
The results for two randomly-selected initial state configurations are
shown in (a) and (b).
In each sub-figure, curves from outside to inside are for $N=50,~100,~250$,
respectively.
The super-period is not affected by chain length, the peak value of the super modulation is kept at unity.
}
\label {Sup_Peak}
\end{figure}

\begin{figure}[htbp]
\begin{center}
    \includegraphics[width=3in]{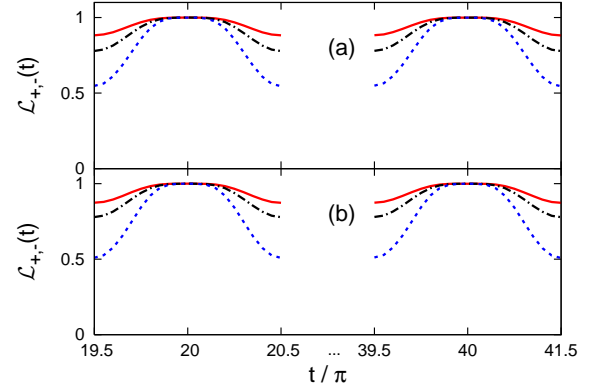}
\end{center}
 \caption[Sup_Peak_Add]
{ (color online).
Modulation on the periodic evolution with $A^{dd}=-2B^{dd}$ for $\xi=40$.
The first two peaks of the super modulation at $t=T_{sup}$ and $t=2T_{sup}$
are shown in detail.
The results for two randomly-selected initial state configurations are
shown in (a) and (b).
In each sub-figure, curves from outside to inside are for $N=50,~100,~250$,
respectively.
The super-period is not affected by chain length, the peak value of the super modulation is kept at unity.
}
\label {Sup_Peak_Add}
\end{figure}

We further investigate the bath size dependence of periodic evolutions
with $N=50,100,\ldots 250$.
We choose relatively large $\xi$ to filter the effect of the super modulation
and find that, not only the basic period $\pi$ but also the peak value
of the oscillating coherence is kept unchanged, see \fig{Effect_L_Pi},
i.e., the $T_{dd}$ periodicity is robust against the
extending of chain length.
On the other hand, the dip of each oscillation period moves down gradually,
and the width of the peak keeps shrinking as the chain length
increases, see \fig{Effect_L_Pi}.

We further attend to the long-time behavior of the periodic evolution.
As shown in \fig{Effect_L}, the super modulation periodicity
still keeps unchanged with the increase of chain length, and the peak
value of the coherence evolution, which occurs at the nodes of the
super-modulation, maintains at unity, see \fig{Sup_Peak}.

We now take the diagonal interaction $A^{dd}=-2B^{dd}$ into consideration.
As the chain size increases, although the
amplitude of the modulation increases as shown in
\fig{Effect_L_Add}, the super-period keeps invariant, and the peak
value of the super-modulated coherence evolution maintains unchanged
as in the case with $A^{dd}=0$, see \fig{Sup_Peak_Add}. This actually provides
a strong numerical evidence that the periodic
evolution even with $A^{dd}=-2B^{dd}$ included is stable.

Above all, the investigation on the chain-length dependence provides
clear evidence that the calculated coherence evolutions in the main
text is intrinsic.


\end{document}